\documentclass[12pt]{iopart}

\usepackage[utf8]{inputenc}
\usepackage[margin=1.0in]{geometry}
\usepackage{graphicx,amstext}
\usepackage{amsfonts}
\usepackage{amssymb}

\usepackage{hyperref}
\usepackage{url}
\usepackage{cite}
\usepackage{color}
\usepackage{ifthen}
\newboolean{editorial}
\setboolean{editorial}{true}
\newcommand{\editorial}[2]{\ifthenelse{\boolean{editorial}}{\textcolor{red}{[\textsf{\textbf{{#1}}}: }\textcolor{blue}{\textsf{{#2}}}\textcolor{red}{]}}{}}

\begin{document}

\title{A cosmologically motivated reference formulation of numerical relativity}

\author{John T. Giblin, Jr${}^{1,2}$}
\author{James B. Mertens${}^{1}$}
\author{Glenn D. Starkman${}^{1}$}

\address{${}^1$CERCA/ISO, Department of Physics, Case Western Reserve University, 10900 Euclid Avenue, Cleveland, OH 44106, USA}
\address{${}^2$Department of Physics, Kenyon College, 201 N College Rd, Gambier, OH 43022, USA}

\ead{james.mertens@case.edu}

\begin{abstract}
The application of numerical relativity to cosmological spacetimes is
providing new insights into the behavior of Einstein's equations, beyond
common approximations. In order for simulations to be performed as accurately
and efficiently as possible, we investigate a novel formulation of Einstein's
equations. This formulation evolves differences from a ``reference'' solution
describing the dominant behavior of the spacetime, which mitigates error due
to both truncation and approximate finite difference calculations. We find
that the error in solutions obtained using the reference formulation can be
smaller by an order of magnitude or more, with the level of improvement
depending on how well the reference solution approximates the exact solution.
\end{abstract}

\submitto{\CQG}
\maketitle

\section{Introduction}

Standard cosmological models contain a number of assumptions, often necessary
in order to make these models tractable. One common assumption is that the
gravitational physics can be simplified, as cosmological systems are often
well-described by linearized gravity or Newtonian physics on an expanding
homogeneous background. Yet, such assumptions have not been thoroughly
tested and compared to predictions from the full theory of general relativity,
leaving open the possibility that these simplifications result in an inaccurate
or biased view of our Universe. Already, observations interpreted utilizing
such assumptions have found moderate inconsistencies among inferred cosmological
parameters \cite{Verde:2013wza,Bernal:2016gxb}. Exploring the validity of
gravitational approximations with sufficient accuracy will require
a careful, systematic analysis of cosmological models in a fully general
relativistic setting.

A number of studies have speculated that quantities computed using
approximate gravitational models in a cosmological context require
percent-level corrections, with even larger corrections possible, depending
on the quantity of interest \cite{Clarkson:2011zq,ellis2012relativistic,Fleury:2016fda}.
Properly characterizing such phenomena will require modeling spacetimes
with at least this accuracy and precision---a formidable task, even in the
context of Newtonian gravity \cite{Schneider:2015yka}. In order to explore
ideas such as this, numerical relativity is being employed in a cosmological
setting with increasing frequency
\cite{Bentivegna:2013jta,Yoo:2014boa,Giblin:2015vwq,Bentivegna:2015flc,Giblin:2016mjp,Macpherson:2016ict,Daverio:2016hqi}.

As a unique method for improving the accuracy of cosmological simulations in
full general relativity, we explore a cosmologically-motivated ``reference
formulation'' in which the fully relativistic differences from an approximate,
semi-analytic background solution are computed. It should be noted that this
is simply a reformulation of Einstein's equations, not an application of
perturbation theory, nor an approximation of Einstein's equations.
The potential benefits of this method are twofold: first, resolving
gravitational physics that could not otherwise be resolved due to roundoff
error, and second, minimizing error contributions due to finite differencing.

We study the accuracy of this method primarily in a cosmological setting,
evolving a universe filled with a pressureless ``dust'' perfect fluid. We
find that the method fares well when the reference solution is a good
description of the spacetime, reducing solution error by an
order-of-magnitude or more, with this advantage decreasing as the
spacetime evolves further away from the reference solution.

In Sec.~\ref{sec:formulation} we develop the reference formulation with
cosmological applications in mind, first presenting details of the standard
BSSNOK system in Sec.~\ref{sec:bssn}, then extending the formulation to
account for the reference solution in Sec.~\ref{sec:reference}. In
Sec.~\ref{sec:ICs} we describe initial conditions for a cosmologically-motivated
test of the reference formulation, and explore the behavior of simulation
error in the reference formulation compared to the standard BSSNOK
formulation in Sec.~\ref{sec:results}.

\section{Constructing a reference formulation}
\label{sec:formulation}

\subsection{The BSSNOK Formulation}
\label{sec:bssn}

Over the past decade, a variety of formulations of general relativity
have been developed for numerical integration, and have demonstrated
their ability to accurately model strongly gravitating systems with
numerically stability. Perhaps the most commonly used 
is the BSSNOK formulation, a conformal 3+1 decomposition of the Einstein
field equations \cite{Nakamura:1987zz,Shibata:1995we,Baumgarte:1998te}.
In this decomposition, the line element is given by
\begin{equation}
ds^{2}=-\alpha^{2}dt^{2}+\gamma_{ij}\left(dx^{i}+\beta^{i}dt\right)\left(dx^{j}+\beta^{j}dt\right)\,,
\end{equation}
where $\alpha$ and $\beta^{i}$ are gauge variables, respectively
known as the lapse and shift. The spatial metric, $\gamma_{ij}$,
is decomposed into a conformal metric, $\bar{\gamma}_{ij}$, and
conformal factor, $\phi$, as $\gamma_{ij}=e^{4\phi}\bar{\gamma}_{ij}$,
where the conformal metric has unit determinant and $\det\gamma_{ij} = e^{12\phi}$.

When written in terms of these variables, the Einstein field equations
are equivalent to the standard BSSNOK equations,
\begin{eqnarray}
\partial_{t}\phi = & -\frac{1}{6}\alpha K+\beta^{i}\partial_{i}\phi+\frac{1}{6}\partial_{i}\beta^{i}\\
\partial_{t}\bar{\gamma}_{ij} = & -2\alpha\bar{A}_{ij}+\beta^{k}\partial_{k}\bar{\gamma}_{ij}+\bar{\gamma}_{ik}\partial_{j}\beta^{k}+\bar{\gamma}_{kj}\partial_{i}\beta^{k}-\frac{2}{3}\bar{\gamma}_{ij}\partial_{k}\beta^{k}\\
\partial_{t}K = & -\gamma^{ij}D_{j}D_{i}\alpha+\alpha(\bar{A}_{ij}\bar{A}^{ij}+\frac{1}{3}K^{2})+4\pi\alpha(\rho+S)+\beta^{i}\partial_{i}K\\
\partial_{t}\bar{A}_{ij} = & e^{-4\phi}(-(D_{i}D_{j}\alpha)+\alpha(R_{ij}-8\pi S_{ij}))^{TF}+\alpha(K\bar{A}_{ij}-2\bar{A}_{il}\bar{A}_{j}^{l})\nonumber \\
 & +\beta^{k}\partial_{k}\bar{A}_{ij}+\bar{A}_{ik}\partial_{j}\beta^{k}+\bar{A}_{kj}\partial_{i}\beta^{k}-\frac{2}{3}\bar{A}_{ij}\partial_{k}\beta^{k}\,.
\end{eqnarray}
Here, the stress-energy tensor has been projected onto the spatial
hypersurfaces of the decomposition, resulting in source terms for the BSSNOK metric
fields, $\rho$, $S_{ij}$, and $S=\gamma^{ij}S_{ij}$. The variable $K$
is the trace of the extrinsic curvature $K_{ij}$, and $\bar{A}_{ij}$ is
the conformally related trace-free part of the extrinsic curvature,
\begin{equation}
e^{4\phi}\bar{A}_{ij}=K_{ij}-\frac{1}{3}\gamma_{ij}K\,.
\end{equation}
In general, quantities with bars are raised, lowered, and computed
using the conformal metric $\bar{\gamma}_{ij}$, and unbarred quantities
with the full 3-metric, $\gamma_{ij}$.

In the BSSNOK formulation, additional auxiliary variables are introduced
to improve stability of the system. A contraction of a Christoffel
symbol of the conformal metric is evolved,
\begin{eqnarray}
\partial_{t}\bar{\Gamma}^{i} = & -2\bar{A}^{ij}\partial_{j}\alpha+2\alpha\left(\bar{\Gamma}_{jk}^{i}\bar{A}^{jk}-\frac{2}{3}\bar{\gamma}^{ij}\partial_{j}K-8\pi\bar{\gamma}^{ij}S_{j}+6\bar{A}^{ij}\partial_{j}\phi\right) \nonumber \\
 & +\beta^{j}\partial_{j}\bar{\Gamma}^{i}-\bar{\Gamma}^{j}\partial_{j}\beta^{i}+\frac{2}{3}\bar{\Gamma}^{i}\partial_{j}\beta^{j}+\frac{1}{3}\bar{\gamma}^{li}\partial_{l}\partial_{j}\beta^{j}+\bar{\gamma}^{lj}\partial_{l}\partial_{j}\beta^{i}\,,
\end{eqnarray}
where $\bar{\Gamma}^{i} = \bar{\gamma}^{jk}\bar{\Gamma}_{jk}^{i}$. A
more comprehensive introduction to this formulation of numerical relativity,
and the motivation behind this scheme, can be found in textbooks such
as \cite{BaumgarteShapiroBook} and \cite{Alcubierre:1138167}.

\subsection{A Cosmologically Motivated Reference Formulation}
\label{sec:reference}

For many spacetimes, there are known solutions of the Einstein (and thus
BSSNOK) equations that describe the spacetime, both approximate and exact,
with deviations from such solutions expected to be small. The focus of this
paper is to examine the behavior of solution error of simulations in a
cosmological setting obtained by evolving deviations from dominant reference
functions, rather than directly using the BSSNOK equations themselves. These
reference functions need not be actual solutions to the Einstein field
equations, and need not be known analytically. The goal will be to use such
reference functions to reduce the susceptibility of calculations to truncation
error or to errors from taking finite differences.

Similar ideas have been explored before \cite{Brown:2009dd}, particularly
in the context of coordinate systems with singularities such as spherical
polar coordinates \cite{Baumgarte:2015dya}. However, to our knowledge, more
general reference functions have not been used.

In a matter-dominated cosmology in geodesic slicing, the dominant solution
can be described purely by several variables of interest. We will be evolving
differences between these variables and the full BSSNOK variables. We denote
difference variables by $\Delta$'s, and define them in terms of reference
functions (hatted) and standard BSSNOK metric variables,
\begin{eqnarray}
\Delta\bar{\gamma}_{ij} & \equiv\bar{\gamma}_{ij}-\delta_{ij}\\
\Delta\bar{\gamma}^{ij} & \equiv\bar{\gamma}^{ij}-\delta^{ij}\\
\Delta\phi & \equiv\phi-\hat{\phi}\\
\Delta K & \equiv K-\hat{K}\\
\Delta \Gamma^i & \equiv \Gamma^i - \hat{\Gamma}^i\\
\Delta\rho & \equiv\rho-\hat{\rho}\,.
\end{eqnarray}
with the remaining metric variable $\bar{A}_{ij}$ being zero. Additional details
regarding computing metric components can be found in \ref{sec:ref_algebra}. Once
expressions for the evolution of the reference variables have been specified,
they can be subtracted from the BSSNOK equations in order to form evolution
equations for the difference variables. Because the BSSNOK equations involve
only differentiation and multiplication, the leading-order contributions to the
BSSNOK equations from the reference solution can be canceled.

Performing this procedure for the differenced conformal metric $\Delta\bar{\gamma}_{ij}$
is straightforward, as the reference metric is taken to be $\delta_{ij}$,
which does not change. The right-hand side of the evolution equations
for the difference metric is therefore identical to the BSSNOK equations,
\begin{equation}
\partial_{t}\Delta\bar{\gamma}_{ij}\equiv\partial_{t}\bar{\gamma}_{ij}\,.
\end{equation}

The remaining variables are chosen to be an approximate solution to
the Einstein field equations. For cosmological systems, an approximate
solution for a pressureless perfect fluid with zero velocity in geodesic
slicing may be found by neglecting $\mathcal{O}(\bar{A}_{ij}^{2})$
terms in the evolution equations for $K$ and $\phi$, as in \cite{Giblin:2015vwq}.
Evolution equations for the conformal factor and the trace of the extrinsic
curvature therefore obey a set of coupled ODEs at each point,
\begin{eqnarray}
\partial_{t}\hat{\phi} & = & -\frac{1}{6}\alpha\hat{K}\,,\\
\partial_{t}\hat{K} & = & \frac{1}{3}\alpha\hat{K}^{2}+4\pi\alpha\hat{\rho}\,.
\end{eqnarray}
These equations describe each location in the universe as obeying
a locally-FLRW equation. In principle, because fluctuations around
this solution are small on the scales of interest (as shown in \cite{Giblin:2015vwq}),
the precision with which algebraic operations can be performed is increased.

Deviations of the extrinsic curvature from the reference solution
will subsequently be sourced only by the neglected $\bar{A}_{ij}\bar{A}^{ij}$
term. This variable, in turn, is both self-sourced, and sourced by the
(trace-free) Ricci tensor. The dominant contribution to the Ricci tensor
comes from derivatives of the conformal factor; thus, accurately determining
these derivatives is important.

The dominant source of error comes from discretization, or from computing
derivatives using finite-difference stencils. To minimize the contribution
of this error, we also evolve the gradients of the reference variables locally,
\begin{eqnarray}
\partial_{t}\hat{\phi}_{,i} & = & -\frac{1}{6}\alpha\hat{K}_{,i}\\
\partial_{t}\hat{K}_{,i} & = & \frac{2}{3}\alpha\hat{K}\hat{K}_{,i}+4\pi\alpha\hat{\rho}_{,i}\\
\partial_{t}\hat{\phi}_{,ij} & = & -\frac{1}{6}\alpha\hat{K}_{,ij}\\
\partial_{t}\hat{K}_{,ij} & = & \frac{2}{3}\alpha\hat{K}_{,i}\hat{K}_{,j}+\frac{2}{3}\alpha\hat{K}\hat{K}_{,ij}+4\pi\alpha\hat{\rho}_{,ij}\,,
\end{eqnarray}
from which gradients of BSSNOK fields can be constructed using,
eg., $\partial_{i}\phi=\hat{\phi}_{,i}+\partial_{i}\Delta\phi$.
Additional reference variables can be defined for each of the remaining
metric variables. The most straightforward of these to write is an equation
for the conformal Christoffel and its derivative,
\begin{eqnarray}
\partial_{t}\hat{\bar{\Gamma}}^{i} & = & -\frac{4}{3}\alpha\delta^{ij}\hat{K}_{,j}\,,\\
\partial_{t}\hat{\bar{\Gamma}}_{,k}^{i} & = & -\frac{4}{3}\alpha\delta^{ij}\hat{K}_{,jk}\,.
\end{eqnarray}

In principle, an equation for the conformal 3-metric (beyond the flat-space
contribution) and its time derivative $\bar{A}_{ij}$ could be written.
However, the evolution equation for $\bar{A}_{ij}$ is sourced by
the Ricci tensor, which requires evolving derivatives of the conformal
factor and conformal metric, which in turn would require evolving
derivatives of ever-increasing order. This procedure could nevertheless
be performed and truncated at some order (even at lowest order, i.e.
$\bar{A}_{ij}$ could be sourced purely by the Ricci tensor assuming
$\bar{\gamma}_{ij}=\delta_{ij}$); however, we leave such an idea for
future work.

Finally, the system is closed by evolving a pressureless perfect-fluid
stress-energy source mimicking a dark matter component. Such a fluid at rest
in geodesic slicing obeys a simple conservation law,
\begin{equation}
\partial_{t}D=\partial_{t}\hat{D}=0\,,
\end{equation}
where the reference and standard density variables are given by
\begin{eqnarray}
\hat{D} & \equiv & \alpha e^{6\hat{\phi}}\hat{\rho}\\
D & \equiv & \alpha e^{6\phi}\rho\,.
\end{eqnarray}
The gradients of the source, $\hat{\rho}_{,i}$ and $\hat{\rho}_{,ij}$ can
be derived from these expressions,
\begin{eqnarray}
\hat{\rho}_{,i} & = & -6\hat{\rho}\hat{\phi}_{,i}+\frac{1}{\alpha}e^{-6\hat{\phi}}\hat{D}_{,i}\\
\hat{\rho}_{,ij} & = & -3\left(\hat{\rho}_{,j}\hat{\phi}_{,i}+\hat{\rho}_{,i}\hat{\phi}_{,j}+2\hat{\rho}\hat{\phi}_{,ij}\right)
 +\frac{-3}{\alpha}e^{-6\hat{\phi}}\left(\hat{D}_{,i}\hat{\phi}_{,j}+\hat{D}_{,j}\hat{\phi}_{,i}-\frac{1}{3}\hat{D}_{,ij}\right) \label{eq:der2rho}\,,
\end{eqnarray}
where the right-hand side of Eq.~\ref{eq:der2rho} has additionally been
symmetrized in the $i$ and $j$ indices.

Subtracting the equations of motion from the BSSNOK equations yields
a set of equations for the corresponding difference variables,
\begin{equation}
\begin{array}{lll}
\partial_{t}\Delta\phi & = & \partial_{t}\phi-\partial_{t}\hat{\phi}\\
 & = & -\frac{1}{6}\alpha\Delta K+\beta^{i}\partial_{i}\phi+\frac{1}{6}\partial_{i}\beta^{i}
\end{array}
\end{equation}
\begin{equation}
\begin{array}{lll}
 \partial_{t}\Delta K & = & \partial_{t}K-\partial_{t}\hat{K} \\
  & = & -\gamma^{ij}D_{j}D_{i}\alpha+\alpha\bar{A}_{ij}\bar{A}^{ij}+\frac{1}{3}\alpha\left(\Delta K+2\hat{K}\right)\Delta K\\
 & & +4\pi\alpha\Delta\rho+4\pi\alpha S+\beta^{i}\partial_{i}K
 \end{array}
\end{equation}
\begin{equation}
\begin{array}{lll}
\partial_{t}\Delta\bar{\gamma}_{ij} & = & \partial_{t}\bar{\gamma}_{ij} \\
 & = & -2\alpha\bar{A}_{ij}+\beta^{k}\partial_{k}\bar{\gamma}_{ij}+\bar{\gamma}_{ik}\partial_{j}\beta^{k}+\bar{\gamma}_{kj}\partial_{i}\beta^{k}-\frac{2}{3}\bar{\gamma}_{ij}\partial_{k}\beta^{k}
\end{array}
\end{equation}
\begin{equation}
\begin{array}{lll}
\partial_{t}\Delta\hat{\bar{\Gamma}}^{i} & = & \partial_{t}\bar{\Gamma}^{i}-\partial_{t}\hat{\bar{\Gamma}}^{i} \\
& = & -2\bar{A}^{ij}\partial_{j}\alpha+2\alpha \bar{\Gamma}_{jk}^{i}\bar{A}^{jk} \\
 & & -\frac{4\alpha}{3}\left(\bar{\gamma}^{ij}\partial_{j}\Delta K+\Delta\bar{\gamma}^{ij}\partial_{j}\hat{K} - 8\pi\bar{\gamma}^{ij}S_{j}+6\bar{A}^{ij}\partial_{j}\phi\right) \\
 & & +\beta^{j}\partial_{j}\bar{\Gamma}^{i}-\bar{\Gamma}^{j}\partial_{j}\beta^{i}+\frac{2}{3}\bar{\Gamma}^{i}\partial_{j}\beta^{j}+\frac{1}{3}\bar{\gamma}^{li}\partial_{l}\partial_{j}\beta^{j}+\bar{\gamma}^{lj}\partial_{l}\partial_{j}\beta^{i}\,.
\end{array}
\end{equation}
An important point here is that the equations for the difference variables
contain no zeroth-order components, and the dominant contribution to gradients
can be computed using reference values so long as the reference solution remains
a good approximate description of the spacetime.

\section{Setting initial conditions}
\label{sec:ICs}

In order to specify initial conditions, we require field configurations
that satisfy both the Hamiltonian and momentum constraint equations,
\begin{eqnarray}
0 &=  \mathcal{H}  &= R + K^{2}-K_{ij}K^{ij} - 16\pi\rho\\
0 &=  \mathcal{M}^i  &=  D_{j}(K^{ij}-\gamma^{ij}K) - 8\pi S^{i}\,.
\end{eqnarray}
As we are interested in comparing the accuracy of evolution between methods,
setting analytic initial conditions is desirable. We simplify the constraint
equations by setting the trace-free part of the extrinsic curvature to zero,
$\bar{A}_{ij} = 0$, and choosing the metric to be conformally flat,
$\bar{\gamma}_{ij} = \delta_{ij}$. We can then obtain a solution by specifying
the metric variables $\phi(\vec{x})$ and $K={\rm const}$, letting
$\rho=\rho_{K}+\rho_{\phi}$, and algebraically solving the Hamiltonian constraint
equation for the density variables,
\begin{eqnarray}
K^{2} & = & 24\pi\rho_{K}\,,\\
\label{eq:phi_ics} \nabla^{2}e^{\phi} & = & -2\pi e^{5\phi}\rho_{\phi}\,.
\end{eqnarray}

This decomposition provides us with the physical metric variables, but a
further choice is required in order to determine initial conditions for
the reference variables. A straightforward choice is to simply use an FLRW
solution close to the above initial conditions. However, this will only
reduce roundoff error, not finite differencing error. Thus we instead choose
to place all fluctuations in the reference variables themselves,
\begin{eqnarray}
\hat{\rho} & = & \rho\\
\hat{\phi} & = & \phi\\
\hat{K} & = & K
\end{eqnarray}
so that the difference variables are all initially zero.

Choosing the conformal density $D$ to be entirely in the reference
solution, $D=\hat{D}$, also allows the density reference variable to
be computed using this variable,
\begin{equation}
\Delta\rho=\hat{\rho} \, \left[e^{-6\Delta\phi} -1\right]\,,
\end{equation}
where the expression in brackets can be evaluated using a function
designed to be accurate for small arguments\footnote{For example, using the standard \textsc{C} function \texttt{expm1}.}.

For the purposes of this work, we examine two solutions in a periodic
spacetime: one containing a single wavelength in the $x$-direction, and
one containing a 3-dimensional solution $\phi$-mode in each direction.
For these solutions, the conformal factor on the initial slice is respectively
chosen to be
\begin{eqnarray}
\label{eq:phi_1d} \phi_{1D} & = & A \sin(2\pi x + \varphi_x)\\
\label{eq:phi_3d} \phi_{3D} & = & A \sin(2\pi x + \varphi_x) \sin(2\pi y + \varphi_y) \sin(2\pi z + \varphi_z)
\end{eqnarray}
for some amplitude $A$ that ensures the density is positive everywhere,
and arbitrary phases $\varphi_i$. Fluctuations in the fluid density are then
reconstructed using Eq.~\ref{eq:phi_ics}.

The metric and matter variables can now be fully determined. We choose metric
variables in terms of the Hubble scale on the initial slice, $H_I^{-1}$, so that
\begin{equation}
\rho_K = \frac{3}{8\pi} H_I^2
\end{equation}
and the trace of the extrinsic curvature
is $K = -3 H_I$. The physical simulation volume is chosen to be
$L^3 = (H_I^{-1} / 2)^3$.

\section{Results}
\label{sec:results}

The main results we present are from simulations in which the sinusoidal metric
fluctuations presented in Sec.~\ref{sec:ICs} are resolved by as few points as is
necessary to obtain results. For these simulations, we demonstrate the
ability of the reference formulation to reduce error in the simulation,
and study the behavior as both resolution and finite-difference-method
order are increased, and as the solution evolves away from the analytic solution.

Although these tests demonstrate the ability of reference formulations to reduce
error, they do not demonstrate the ability of the method to resolve fluctuations
on a dominant background spacetime beyond the level of standard formulations.
Therefore, in \ref{sec:awa_lin_wave} we present additional runs of a linearized
gravitational wave propagating through a flat spacetime.

As finite differences must still be computed in the reference formulation, the
method order and convergence rate should remain similar to that of a standard,
non-reference formulation. However, the amplitude of fluctuations around the
analytic solution are expected to be smaller in the reference formulation, 
and to an extent smoother, leading to smaller errors when finite differences must
be computed.

\begin{figure*}[t]
  \begin{flushright}
    \includegraphics[width=0.84\textwidth]{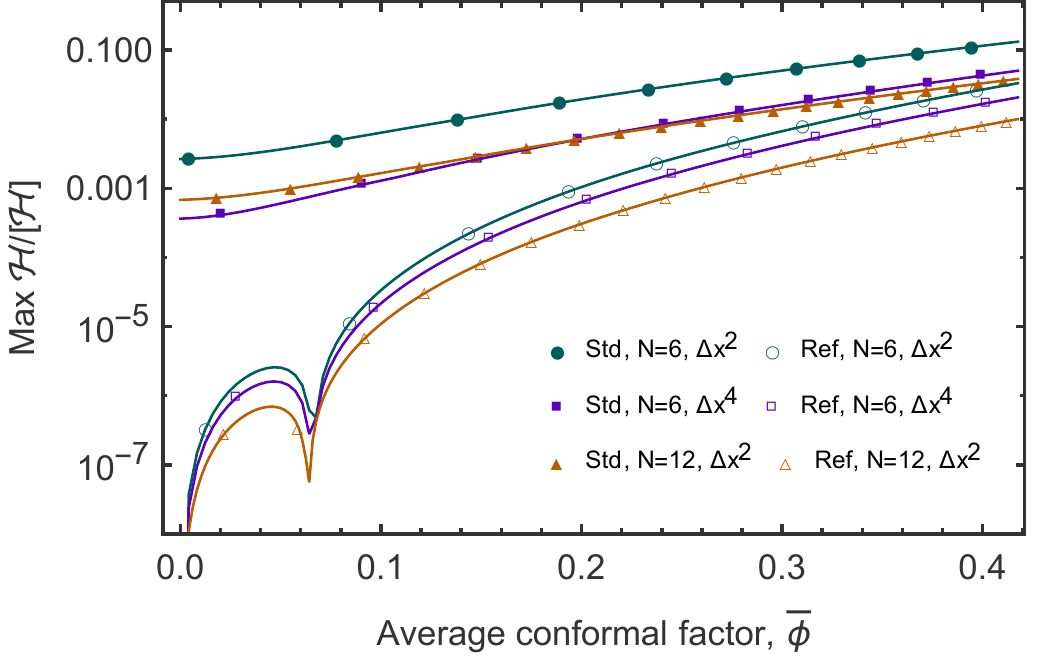}
  \end{flushright}
  \caption{\label{H_viol}
  Hamiltonian constraint violation vs the volume-averaged conformal factor (roughly, half the number of
  cosmological e-folds of expansion undergone) for various 1-dimensional runs. Run parameters are
  indicated by the legend, including resolution ($N$ being the number of points in the $x$-direction, with
  a minimal number of points in other directions due to the symmetry of the simulation),
  finite-difference stencil order (accurate to $\mathcal{O}(\Delta x^p)$), and method type being either the reference
  formulation (`Ref', open shapes) or the standard BSSNOK formulation (`Std', filled shapes). In general,
  the amount of constraint violation is found to be smaller using the reference formulation. 
  }
\end{figure*}

For cosmological runs, there are a number of ways to quantify the error
of the system. In Figure~\ref{H_viol}, we present the amplitude of
constraint violation in standard and reference formulations, for a simulation
using Eq.~\ref{eq:phi_1d} to set initial conditions. These
are plotted against the volume-weighted average conformal factor of the spacetime, 
which increases monotonically with time, and corresponds to roughly half the
number of e-folds of cosmological expansion the simulation has undergone.

The amplitude of fluctuations of the conformal factor is $A = 2 \times 10^{-4}$,
resulting in a conformal standard deviation of the density
$\sigma_{\rho} / \bar{\rho} \sim 0.05$ on the initial slice, and
a minimum and maximum overdensity $\delta_{\rho} / \bar{\rho} \sim 0.1$.
We use a very small timestep, $\Delta t = 10^{-4} \Delta x$, to ensure
time-integration convergence of solutions so that a meaningful comparison of
error arising from discretization effects can be made. The simulations
are then run until the worst-case surpasses $\sim 10\%$
Hamiltonian constraint violation amplitude relative to the energy scale of
the problem, or the maximum $\mathcal{H} / [\mathcal{H}] > 0.01$ for
\begin{equation}
[\mathcal{H}] = \frac{e^{5\phi}}{8}\sqrt{\bar{R}^2 + \left(\frac{2}{3}K^{2}\right)^2 + (A_{ij}A^{ij})^2 + (16\pi\rho)^2}\,.
\end{equation}
Runs are performed using a minimal resolution, $N=6$ in the direction
containing fluctuations, with as few points as necessary to compute finite
differences in other directions. Finite differences are also computed with
minimal accuracy, with error of order $\mathcal{O}(\Delta x^2)$, and compared
to an $\mathcal{O}(\Delta x^4)$ method.

From Fig.~\ref{H_viol}, we observe that the amplitude of measured constraint
violation is smaller when taking advantage of the reference solution. The
benefit of the reference formulation is most pronounced at early simulation
times when the solution is closest to the reference solution. As the simulation
evolves, deviating further from the reference solution, the relative benefit of
the formulation decreases.

Beyond the improvement seen when increasing resolution or method order, or when
using the reference formulation, there are additional noteworthy features. First,
despite the use of analytic initial conditions, runs using the standard BSSNOK
formulation mis-infer the amount of constraint violation present due to finite
differencing error, resulting in a non-zero measurement of constraint violation
on the initial surface. The reference formulation is, at least initially, able
to compute the amount of constraint violation to within machine precision.
Although the amount of error in the simulations does grow, the error demonstrates
appropriate convergence for both formulations.

\begin{figure*}[t]
  \begin{flushright}
    \includegraphics[width=0.84\textwidth]{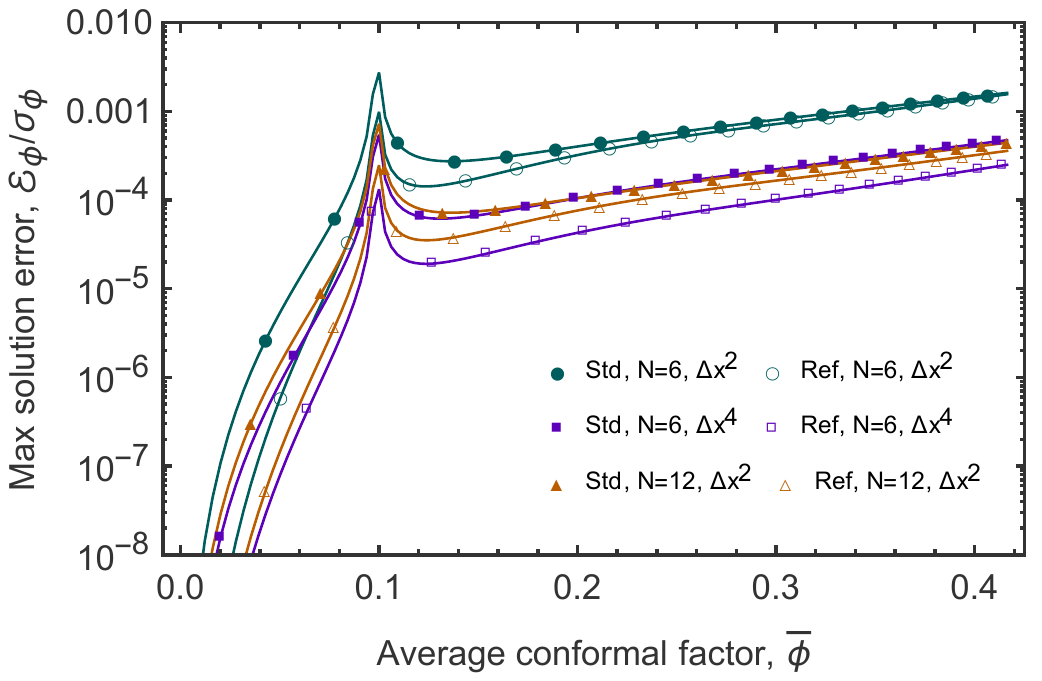}
  \end{flushright}
  \caption{\label{soln_error}
  Solution error relative to an $N = 120$ simulation as a function of the average
  conformal factor, $\bar{\phi}$ (a stand-in for time). Simulations performed
  using the reference formulation are seen to be more accurate, especially at
  early times when the reference solution is a good approximation.
  }
\end{figure*}

As computing the amplitude of constraint violation itself requires evaluating
finite differences, it suffers from precisely the type of error we are trying to
mitigate. Therefore, looking solely at the constraint violation does not
provide the best measure of error. Because we are running at a
low resolution, we can simply compare to higher resolution runs to obtain a more
meaningful comparison. A solution obtained with $N = 120$ using an 8th-order
finite-difference method is taken to be the true solution, and deviations
from this solution are plotted in Figure~\ref{soln_error}.

As with the constraint violation measure, a relative benefit can be seen when
using the reference formulation. For these runs, the maximum solution error of
the local conformal factor $\varepsilon_\phi \equiv |\phi - \phi_{\rm true}|$
is plotted relative to the RMS amplitude of the
fluctuations, $\sigma_\phi$, so as to emphasize the accuracy with which these
fluctuations have been computed rather than the accuracy with which the dominant
background cosmology has been computed. Again, the relative benefit is seen to
be largest at early times, when the solution is well-described by the
approximate reference solution. In this case, the formulation offers an
order-of-magnitude benefit, and can be seen to perform better than 
doubling either the resolution or method order. At late times, however, the
relative benefit disappears.

Several final points are of note. First, the reduction in relative error is
seen to be somewhat greater for higher finite difference orders, which we
speculate is due to the increased smoothness of the reference solution.
Second, a spike in the relative error appears around
$\bar{\phi}=0.1$. This is due to a brief drop in the value of $\sigma_\phi$ as
it is driven from initial anti-correlation with density fluctuations towards
becoming correlated (a behavior which can be viewed as a peculiarity of
geodesic slicing). Finally, although not plotted here, the relative benefit at
late times is found to be larger for smaller amplitude fluctuations, i.e. when the
reference solution is a better approximation.

\begin{figure*}[htb]
  \begin{flushright}
    \includegraphics[width=0.84\textwidth]{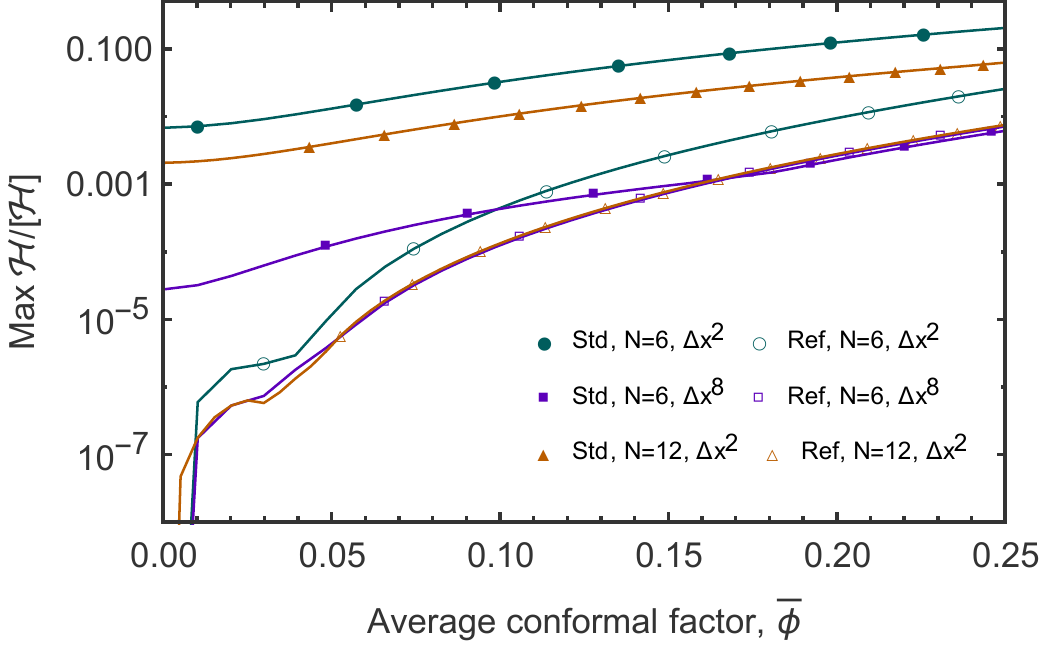}
  \end{flushright}
  \caption{\label{H_viol_3d}
  Hamiltonian constraint violation as a function of the volume-averaged conformal
  factor for various 3-dimensional runs. Run parameters are indicated by the legend,
  as in previous figures. As with the 1-dimensional case, the computed amount of
  constraint violation is smaller when using the reference formulation.
  }
\end{figure*}

\begin{figure*}[htb]
  \begin{flushright}
    \includegraphics[width=0.84\textwidth]{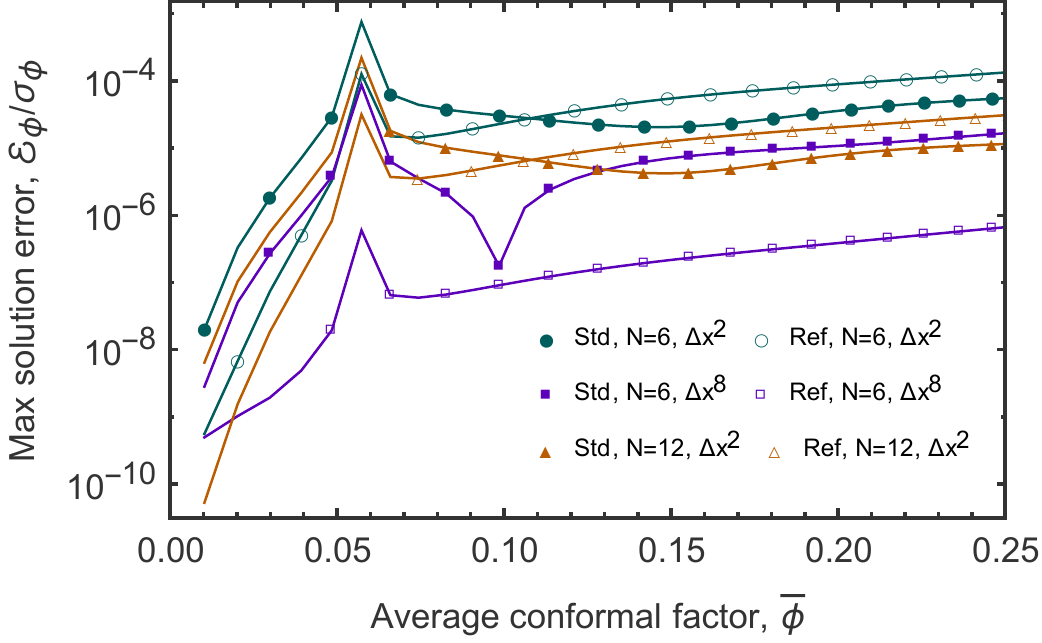}
  \end{flushright}
  \caption{\label{soln_error_3d}
  Solution error as a function of the average conformal factor, $\bar{\phi}$.
  Simulations performed using the reference formulation are found to be more
  accurate at early times, although less accurate at late times when using
  a low-order finite difference method.
  }
\end{figure*}

We additionally show the results of a 3-dimensional simulation with $N^3$
points, with initial conditions described by Eq.~\ref{eq:phi_3d}. These
simulations are again run until thee worst case has surpassed roughly 10\%-level
constraint violation. Results from these runs are plotted in
Figures \ref{H_viol_3d} and \ref{soln_error_3d}. For these runs, we
find similar results when looking at the level of computed constraint
violation: the reference formulation yields a substantially smaller inferred
level of constraint violation. However, again, the solution error should provide
a better measure of simulation accuracy. For a low-order finite difference
method, the advantage of the reference formulation disappears as the evolution
progresses far enough away from the analytic solution. This remains true at higher
resolutions, but is no longer true for a higher order finite difference method,
suggesting that this is due to the numerical solution obtained using the reference
formulation becoming insufficiently smooth at the level of the order of the finite
difference error. However, for higher-order methods where finite differences
are computed more accurately, the reference formulation maintains a significant
advantage for the duration of the simulation. For both methods and for all
choices of numerical and physical parameters, proper numerical convergence is
found.

\section{Discussion}

We have presented results from a cosmological reference formulation
in which the dominant behavior of the spacetime dynamics was well-described by an
approximate reference solution. The formulation was seen to provide more
accurate results when the reference solution was a good description of the spacetime,
and to nevertheless converge even when the reference solution was no longer a good
approximation. The reference formulation was also found to be increasingly
accurate relative to the non-reference formulation as method order increased.

Several improvements to the cosmological reference formulation presented here
can be made that would ostensibly decrease simulation error. The evolution
could use a slicing condition with improved stability properties, such as
Harmonic slicing. In addition to the reference variables shown above, such a
gauge would also require introducing a reference lapse, and spatial derivatives
of the lapse. A second improvement would be to evolve the conformal metric
and extrinsic curvature components sourced by all terms that do not require
computing finite differences.

As we work to simulate cosmological systems in a fully relativistic setting with
increasing realism, obtaining accurate results as efficiently as possible is an
important goal. This formulation has demonstrated the ability to obtain reliable
results and considerably decrease the numerical error without making any
additional approximations.

\section*{Acknowledgments}

We would like to thank Thomas Baumgarte for a series of very valuable
conversation that helped shape this work. JTG is supported by the National
Science Foundation, PHY-1414479; JBM and GDS are supported by a Department of
Energy grant DE-SC0009946 to CWRU. The simulations in this work made use of
hardware provided by the National Science Foundation and the Kenyon College
Department of Physics, and the High Performance Computing Resource in the
Core Facility for Advanced Research Computing at Case Western Reserve
University.

\section*{\refname}
\bibliography{references}

\appendix
\section{Accurate calculation of algebraic quantities}
\label{sec:ref_algebra}

From the definitions in Sec.~\ref{sec:reference}, care should be taken when
raising and lowering indices of difference variables so that they are not
directly raised and lowered using the 3-metric $\bar{\gamma}_{ij}$. Given a
known difference metric $\Delta\bar{\gamma}_{ij}$, the difference of the
inverse can be computed in an algebraic manner by explicitly writing out
matrix components inverting, and subtracting. Noting that both
$\det\bar{\gamma}_{ij}=\det\hat{\bar{\gamma}}_{ij}=1$,
the end result of this operation results in matrix components
\begin{eqnarray}
\fl \Delta\bar{\gamma}^{11} & = & \hat{\bar{\gamma}}_{33}\Delta\bar{\gamma}_{22}-\Delta\bar{\gamma}_{23}(2\hat{\bar{\gamma}}_{23}+\Delta\bar{\gamma}_{23})+(\hat{\bar{\gamma}}_{22}+\Delta\bar{\gamma}_{22})\Delta\bar{\gamma}_{33} \nonumber\\
\fl \Delta\bar{\gamma}^{12} & = & -\hat{\bar{\gamma}}_{33}\Delta\bar{\gamma}_{12}+\hat{\bar{\gamma}}_{23}\Delta\bar{\gamma}_{13}+(\hat{\bar{\gamma}}_{13}+\Delta\bar{\gamma}_{13})\Delta\bar{\gamma}_{23}-(\hat{\bar{\gamma}}_{12}+\Delta\bar{\gamma}_{12})\Delta\bar{\gamma}_{33} \nonumber\\
\fl \Delta\bar{\gamma}^{13} & = & \hat{\bar{\gamma}}_{23}\Delta\bar{\gamma}_{12}-\hat{\bar{\gamma}}_{22}\Delta\bar{\gamma}_{13}-(\hat{\bar{\gamma}}_{13}+\Delta\bar{\gamma}_{13})\Delta\bar{\gamma}_{22}+(\hat{\bar{\gamma}}_{12}+\Delta\bar{\gamma}_{12})\Delta\bar{\gamma}_{23} \nonumber\\
\fl \Delta\bar{\gamma}^{22} & = & \hat{\bar{\gamma}}_{33}\Delta\bar{\gamma}_{11}-\Delta\bar{\gamma}_{13}(2\hat{\bar{\gamma}}_{13}+\Delta\bar{\gamma}_{13})+(\hat{\bar{\gamma}}_{11}+\Delta\bar{\gamma}_{11})\Delta\bar{\gamma}_{33} \nonumber\\
\fl \Delta\bar{\gamma}^{23} & = & -\hat{\bar{\gamma}}_{23}\Delta\bar{\gamma}_{11}+\hat{\bar{\gamma}}_{13}\Delta\bar{\gamma}_{12}+(\hat{\bar{\gamma}}_{12}+\Delta\bar{\gamma}_{12})\Delta\bar{\gamma}_{13}-(\hat{\bar{\gamma}}_{11}+\Delta\bar{\gamma}_{11})\Delta\bar{\gamma}_{23} \nonumber\\
\fl \Delta\bar{\gamma}^{33} & = & \hat{\bar{\gamma}}_{22}\Delta\bar{\gamma}_{11}-\Delta\bar{\gamma}_{12}(2\hat{\bar{\gamma}}_{12}+\Delta\bar{\gamma}_{12})+(\hat{\bar{\gamma}}_{11}+\Delta\bar{\gamma}_{11})\Delta\bar{\gamma}_{22}\,.
\end{eqnarray}
All terms in these equations are, notably, multiples of difference
metric components; thus for small differences from the reference metric,
components of the inverse metric are also small. Conformal Christoffel
symbols are also computed using the difference metric,
\begin{equation}
\bar{\Gamma}_{ijk}=\frac{1}{2}\left(\partial_{j}\Delta\bar{\gamma}_{ik}+\partial_{k}\Delta\bar{\gamma}_{ij}-\partial_{i}\Delta\bar{\gamma}_{jk}\right)\,.
\end{equation}

\section{Linearized Gravitational Wave Test}
\label{sec:awa_lin_wave}

In this section we present results from a small-amplitude, linearized gravitational
wave test around flat space. This is precisely the Apples with Apples linearized
wave test \cite{Babiuc:2007vr}, however the amplitude is substantially reduced in
order to demonstrate the ability of the formulation to resolve small fluctuations
around a dominant background---in this case, simply the flat, Minkowski metric.

The noteworthy feature here is that differences from $\delta_{ij}$ are evolved
directly. For diagonal metric components in particular, this means that the
dominant `1' does not factor into roundoff error. While this particular test
shows that features of such small amplitudes can be resolved, it does not
demonstrate the ability of the method to model nonlinear physics more
accurately, as is the intent behind the main results in Sec.~\ref{sec:results}.

The linearized wave Apples with Apples (AwA) test examines a linearized gravitational
wave solution whose metric is of the form
\begin{equation}
ds^2 = -dt^2 + dx^2 + (1+H)dy^2 + (1-H)dz^2\,,
\end{equation}
with
\begin{equation}
\label{AwA_H}
H = A \sin(2\pi(x-t))\,,
\end{equation}
for a unit length box. The simulation is run for 1000 box-crossing times
and checked against the analytic solution.
The value of $A$ here is usually taken to be $A = 10^{-8}$, so that terms
of order $A^2$ will be at the level of roundoff error, or a part in $10^{16}$
when compared to the dominant `1' of the flat metric. Here we will present
results from a run with $A = 10^{-16}$ so that second-order terms are still
at the level of roundoff. However, they will now be at the level of roundoff
relative to the amplitude of the gravitational wave itself.

\begin{figure*}[htb]
  \centering
    \includegraphics[width=1.0\textwidth]{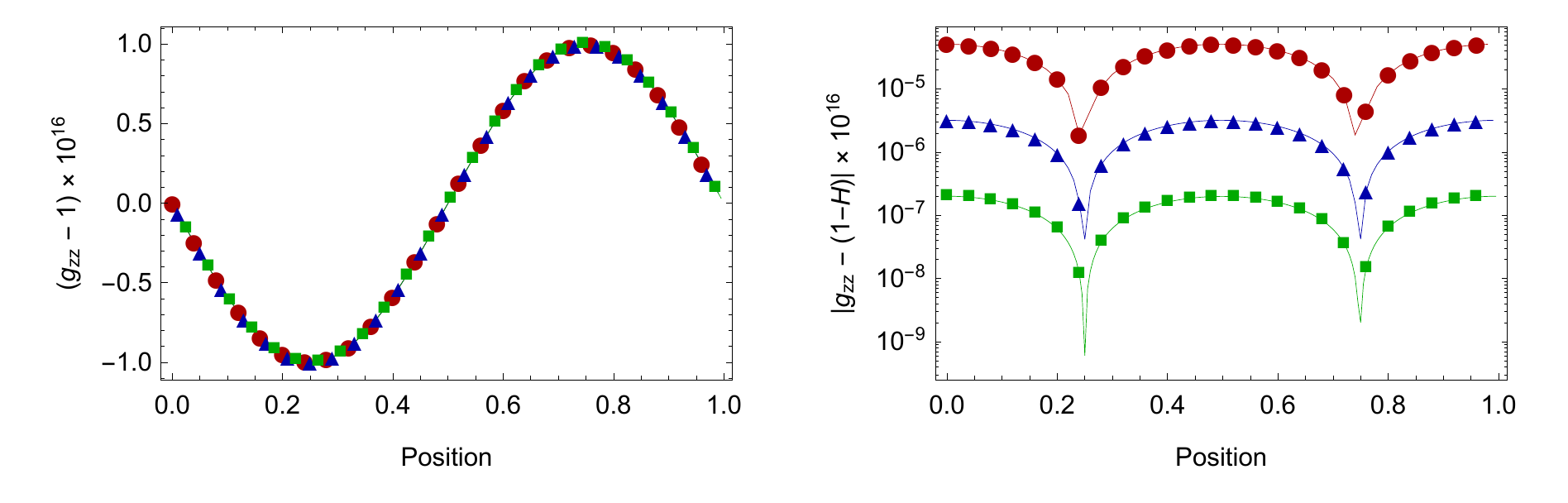}
  \caption{\label{linwave_soln}
  	Results from the AwA linear wave test with a very small wave amplitude.
    Shown is the numerical solution for $g_{zz}$, and the difference of this solution
    from the analytic, linear solution ($1-H$) after 1000 box-crossing times. Good
    agreement with the analytic answer demonstrates the ability of the formulation
    to resolve very small fluctuations around a background spacetime. The test is
    run for 3 resolutions: 50 points (red, circles), 100 points (blue, triangles),
    and 200 points (green, squares).
  }
\end{figure*}

As per test specifications, the simulation was run for 1000 box-crossing
times with a timestep $\Delta t = \Delta x/4$. Results from this test are shown in
Fig.~\ref{linwave_soln}. The observed convergence rate is in agreement with a
method accurate to 4th-order in $\Delta x$, consistent with the dominant simulation
error arising from the RK4 integration scheme used. Finite difference stencils were
computed using an 8th-order scheme.

\end{document}